\documentclass[12pt,thmsa]{article}
\usepackage{sw20aip}

\input{tcilatex}
\begin{document}

\title{Entropy inequalities and Bell inequalities for two-qubit systems}
\author{Emilio Santos \and Departamento de F\'{i}sica. Universidad de Cantabria.
Santander. Spain}
\maketitle

\begin{abstract}
Sufficient conditions for the non-violation of the Bell-CHSH inequalities in
a mixed state of a two-qubit system are: 1$)$ The linear entropy of the
state is not smaller than 1/2, 2) The sum of the conditional linear
entropies is not negative, 3) The von Neumann entropy is not smaller than
0.833, 4) The sum of the conditional von Neumann entropies is not smaller
than 0.280.

\end{abstract}

\section{\protect\smallskip Introduction}

\smallskip As is well known, entangled quantum states give rise to most
counterintuitive features. For instance, in classical physics, as well as in
all other branches of science except quantum mechanics, complete knowledge
of a composite system requires knowledge of every one of its parts. Indeed
this is a common definition of \thinspace ``complete knowledge''. In sharp
contrast, in quantum mechanics if we know that two spin-1/2 particles are in
a state of zero total spin, our knowledge about the spin of the system is
complete, the quantum state being pure, but we have no information at all
about the individual spin of each particle. If the (lack of) information
about a system consisting of two subsystems is formalized by means of the
Shannon entropy, S$_{12},$ and the information about the first (second)
subsystem by S$_{1}$ (S$_{2}$), the above mentioned characteristic of
classical information implies the fulfillement of the entropy inequalities 
\begin{equation}
S_{12}\geq S_{1},S_{12}\geq S_{2},  \label{00}
\end{equation}
which mean that the ignorance about the whole cannot be smaller than the
ignorance about a part. In the rest of this article we shall name $\left( 
\ref{00}\right) $ ``entropy inequalities''.

In quantum mechanics several definitions of entropy have been proposed with
the property that the inequalities analogous to $\left( \ref{00}\right) $
are violated in some cases, e.g. in the singlet spin state mentioned above.
(For a review of quantum entropies see Vedral\cite{vedral} and references
therein.) The most popular quantum entropy is due to von Neumann, but the
most simple one is the so-called linear entropy which, for a system
consisting of two subsystems, is defined as 
\begin{equation}
S_{12}:=Tr\left( \rho \left( 1-\rho \right) \right) \equiv 1-Tr\left( \rho
^{2}\right) ,\;S_{j}:=1-Tr\left( \rho _{j}^{2}\right) ,  \label{03}
\end{equation}
where $\rho $ is the density matrix of the whole system, and $\rho _{j}$ is
the reduced density matrix of subsystem j (j = 1,2). The linear entropy is
usually considered as a standard measure of mixedness of a state. For a
composite system, an interesting property is that the violation of the
inequality $\left( \ref{00}\right) $ is a necessary condition for
entanglement. It holds true in general, not only for two-qubit systems. For
the sake of clarity we give the proof, which is very simple. In fact, a
quantum state of the system is separable if, and only if, its density matrix
may be written in the form 
\begin{equation}
\rho =\sum_{k}w_{k}\rho _{1k}\rho _{2k},w_{k}>0,\sum_{k}w_{k}=1,  \label{02}
\end{equation}
where $\rho _{1k}$ ($\rho _{2k})$ are density matrices of the first (second)
subsystem. If we put $\left( \ref{02}\right) $ into $\left( \ref{03}\right) $
we get, using well known properties of the density matrices, 
\begin{eqnarray*}
S_{12} &=&1-\sum_{k}\sum_{l}w_{k}w_{l}Tr_{1}\left( \rho _{1k}\rho
_{1l}\right) Tr_{2}\left( \rho _{2k}\rho _{2l}\right) \\
&\geq &1-\sum_{k}\sum_{l}w_{k}w_{l}Tr_{1}\left( \rho _{1k}\rho _{1l}\right)
=1-Tr_{1}\left[ \left( \sum w_{k}\rho _{1k}\right) ^{2}\right] =S_{1},
\end{eqnarray*}
where $Tr_{1}(Tr_{2})$ is the trace in the Hilbert space of the first
(second) subsystem, and the inequality derives from $Tr_{2}\left( \rho
_{2k}\rho _{2l}\right) \leq 1.$ This completes the proof that separability
is a sufficient condition for the fulfillement of $\left( \ref{00}\right) $
for quantum linear entropy. Thus the entropy inequalities give a partial
characterization of entanglement, partial because separability, although
sufficient, is not necessary for the fulfillement of the inequalities.

\smallskip Another method for the characterization of non-classical states
of physical systems or, more specifically, to discover whether two distant
physical systems are entangled is the use of Bell\'{}s inequalities. They
have the advantage of connecting quantities which may be measured, at least
in principle. As is well known, the violation of a Bell inequality is a
sufficient condition for entanglement (non-separability). The more general
theoretical question of fully characterizing quantum states compatible with
every Bell inequality is still unsolved\ (it is solved for pure states,
which may violate a Bell inequality if and only if there is entanglement\cite
{selleri} .) In this article we shall consider only the most popular Bell
inequalities, namely the CHSH (Clauser-Horne-Shimony-Holt\cite{CHSH}) or the
equivalent Clauser-Horne\cite{CH} inequalities (for the proof of equivalence
see below, after eq.$\left( \ref{5}\right) $.) Actually there are other Bell
type inequalities, for instance entropic Bell inequalities, which involve
classical entropy, hold true in any classical theory, but may be violated by
quantum mechanics\cite{bra}$^{,}$\cite{cerf} .

In summary, it is known that separability implies the fulfillement of both
Bell\'{}s inequalities and quantum entropy inequalities. Therefore a natural
question is to ask whether the entropy inequalities are stronger or weaker
than the Bell inequalities. That question may have also practical relevance
for the applications of quantum information theory\cite{nielsen} . The
attempt to get an answer is the main motivation for the present paper. The
problem has been already investigated using quantum linear entropy. In fact
it has been shown\cite{HH} that the inequality $\left( \ref{00}\right) $ for
linear entropy is a sufficient condition for all CHSH inequalities. A
slightly more powerful result is also true, namely that 
\begin{equation}
S_{2/1}+S_{1/2}\geq 0,\;\;S_{i/j}:=S_{12}-S_{j},  \label{1}
\end{equation}
where S$_{i/j}$ are called conditional entropies, is sufficient\cite{sf} .
This means that, for quantum linear entropy 
\begin{equation}
separability\Rightarrow \text{ }entropy\text{ }inequalities\Rightarrow Bell%
\text{ }inequality.  \label{implications}
\end{equation}
The specific aim of the present article is to generalize these results
deriving inequalities, possibly weaker than $\left( \ref{00}\right) $,
involving quantum (linear and von Neumann) entropy, which are sufficient for
the non-violation of the CHSH or CH inequalities for a two-qubit system in
any mixed state.

The CHSH inequality is 
\begin{equation}
-2\leq \beta \leq 2,\;\beta \equiv \left\langle a_{1}a_{2}\right\rangle
+\left\langle a_{1}b_{2}\right\rangle +\left\langle b_{1}a_{2}\right\rangle
-\left\langle b_{1}b_{2}\right\rangle ,  \label{5}
\end{equation}
a$_{1}$, b$_{1}$ (a$_{2}$, b$_{2}$) being dichotomic observables, which may
take only the values +1 or -1, for the first (second) qubit and $%
\left\langle x\right\rangle $ means the average of the observable $x$ over
many runs of the same experiment. As is well known the four averages should
be measured in different experiments all of them using the same preparation
for the two-qubit system. I point out that any sufficient condition for the
CHSH inequality is also valid for the Clauser-Horne inequality\cite{CH} 
\begin{equation}
p(A_{1})+p(A_{2})\geq
p(A_{1}A_{2})+p(A_{1}B_{2})+p(B_{1}A_{2})-p(B_{1}B_{2}),  \label{30}
\end{equation}
where A$_{j}$ , B$_{j}$, are observables which may take only the values 1 or
0, and p(X) (or p(XY)) is the probability that X (or both X and Y) takes the
value 1. In fact, it is enough to put 
\[
a_{j}=2A_{j}-1,b_{j}=2B_{j}-1, 
\]
in eq.$\left( \ref{5}\right) $ in order to check that $\beta \leq 2$ implies
eq.$\left( \ref{30}\right) .$

\section{Bell inequalities and linear entropy}

We shall consider quantum observables (Hermitean traceless 2$\times $2
matrices) \{a$_{1}$, b$_{1}$\} for the first qubit and \{a$_{2}$, b$_{2}$\}
for the second, all observables having eigenvalues 1 or -1$.$ We define a
Bell operator\cite{revzen} , $B$, by 
\begin{equation}
B=a_{1}\otimes a_{2}+a_{1}\otimes b_{2}+b_{1}\otimes a_{2}-b_{1}\otimes
b_{2}.  \label{18}
\end{equation}
Hence it is easy to check that (see Appendix) 
\begin{equation}
TrB=0,\;Tr\left( B^{2}\right) =16,  \label{19}
\end{equation}
and that the inequality $\left( \ref{5}\right) $ is violated if, for some
choice of the Bell operator B, $\left| \beta \right| >2$, where 
\begin{equation}
\beta =Tr\left( \rho B\right) ,  \label{20}
\end{equation}
whilst quantum mechanics just predicts $\left| \beta \right| \leq 2\sqrt{2}.$
(Eq.$\left( \ref{20}\right) $ follows from eq.$\left( \ref{5}\right) $ and
the linearity of the trace.)

\smallskip It is the case that not all values of $\beta $ and S$_{12}$ are
compatible. Our problem is to find the region of compatibility, which may be
stated as a mathematical problem as follows. Eq.$\left( \ref{20}\right) $
and the first eq.$\left( \ref{03}\right) $ define a mapping, \textit{M}, of
the set, $\Lambda ,$ of pairs $\left\{ \rho ,B\right\} $ of a density
operator $\rho $ and a Bell operator B, ( both of which may be represented
by 4$\times 4$ matrices) into the set, \textbf{R}$^{2}$, of pairs of real
numbers $\left\{ \beta ,S_{12}\right\} .$ Thus our problem is to find the
the image in the mapping \textit{M}. I shall prove the following:

\begin{theorem}
In a two-qubit system, for any Bell operator, B, and any state, $\rho ,$ the
linear entropy of the state, S$_{12},$\ and the parameter $\beta $\ lie
within a region of the plane $\left\{ \beta ,S_{12}\right\} $ bounded by the
inequalities 
\begin{equation}
S_{12}\geq 0,\;S_{12}\leq \frac{3}{4}-\frac{\beta ^{2}}{16},\;S_{12}\leq 1-%
\frac{\beta ^{2}}{8},  \label{1a}
\end{equation}
The region so defined is minimal in the sense that no smaller region of
compatibility exists.
\end{theorem}

In mathematical terms we may say that the image of the mapping \textit{M} is
the region defined by $\left( \ref{1a}\right) .$

\textit{Proof:} The first inequality is a trivial consequence of the
definition of S$_{12}$ $\left( \ref{3}\right) .$ The second derives from the
obvious inequality 
\[
Tr\left( \rho -\frac{1}{4}I+\eta B\right) ^{2}\geq 0,\eta \in R, 
\]
where \textit{I} is the 4$\times $4 unit matrix. The condition that this
inequality is fulfilled for any real number $\eta $ leads to the second
inequality $\left( \ref{1a}\right) ,$ as may be easily proved taking eqs.$%
\left( \ref{19}\right) $ into account.

The third inequality $\left( \ref{1a}\right) $ is proved as follows. Any
Bell operator, B, has eigenvalues $\left\{ \xi _{j}\right\} $ so that (see
Appendix) 
\begin{equation}
\xi _{2}=\pm \sqrt{8-\xi _{1}^{2}},\;\xi _{3}=-\xi _{2},\;\xi _{4}=-\xi _{1},
\label{1b}
\end{equation}
where $\xi _{1}\in \left[ -2\sqrt{2},2\sqrt{2}\right] .$ Given an arbitrary
pair,\{$\rho ,B$\}$\in \Lambda $ we may write $\rho $ in the basis of the
eigenvectors of B, and label r$_{ij}$ its elements. Without loss of
generality we may assume $\xi _{1},\xi _{2}\geq 0$ (which amounts to
attaching the labels 1 and 2 to the two positive eigenvalues of B). Thus 
\begin{equation}
\beta =\sum_{j}\xi _{j}r_{jj}\leq \beta ^{\prime }\equiv \xi _{1}r_{11}+%
\sqrt{8-\xi _{1}^{2}}r_{22}\leq 2\sqrt{2}\sqrt{r_{11}^{2}+r_{22}^{2}},
\label{2}
\end{equation}
where the first inequality derives from r$_{jj\text{ }}\geq 0$ , the matrix $%
\rho $ being positive, and $\xi _{3},\xi _{4}\leq 0$ (see $\left( \ref{1b}%
\right) ).$ The second inequality $\left( \ref{2}\right) $ holds true
because $\beta ^{\prime }$ as a function of $\xi _{1},$ for fixed positive r$%
_{11}$ and r$_{22},$ possesses an absolute maximum given by the right side$.$
Also 
\[
1-S_{12}\equiv \sum_{i,j}r_{ij}r_{ji}=\sum_{i,j}\left| r_{ij}\right|
^{2}\geq r_{11}^{2}+r_{22}^{2}, 
\]
where the equality derives from the Hermitian character of $\rho $. This
inequality, combined with $\left( \ref{2}\right) $, gives the last
inequality $\left( \ref{1a}\right) $ if $\beta \geq 0$ (for $\beta <0$ the
proof is similar).

Now we shall prove that, for any given values of S$_{12}$ and $\beta $ in
the interior of the region $\left( \ref{1a}\right) ,$ it is possible to find
a Bell operator B and a state $\rho $ leading to these values of S$_{12}$
and $\beta $ via eqs.$\left( \ref{03}\right) $ and $\left( \ref{20}\right) .$
In order to make the proof we shall consider the set, $\Lambda _{C}\subset
\Lambda ,$ of pairs $\left\{ \rho ,B\right\} $ such that the operators $\rho 
$ and $B$ commute. The commutativity implies that the density matrix $\rho $
is diagonal in a basis of eigenvectors of B (these eigenvectors are usually
called Bell states).

We consider firstly a subset, $\Lambda _{1}\subset \Lambda _{C},$ consisting
of pairs such that the elements of the density matrix, $\rho ,$ are, in the
basis of Bell states, 
\[
r_{11}=r_{22}=\frac{1}{4}\left( 1+\alpha \right) ,\;r_{33}=r_{44}=\frac{1}{4}%
\left( 1-\alpha \right) ,\;r_{ij}=0\text{ otherwise,} 
\]
where $\alpha \in \left[ -1,1\right] .$ For this pair $\left\{ \rho
,B\right\} $ eqs.$\left( \ref{03}\right) $ and $\left( \ref{20}\right) $
give 
\[
S_{12}=\frac{3}{4}-\frac{\alpha ^{2}}{4},\text{ }\beta =\frac{\alpha }{2}%
\left( \xi _{1}+\xi _{2}\right) . 
\]
Taking into account the domain of $\alpha $ and the relations $\left( \ref
{1b}\right) $ it is easy to see that the image of the set $\Lambda _{1}$ in
the mapping M possesses a range defined by the inequalities $\left( \ref{1a}%
\right) $ plus S$_{12}\geq 1/2.$

Now we define another subset, $\Lambda _{2}$, consisting of pairs such that
the density matrix, $\rho ,$ written in the basis of eigenvectors of B, has
elements 
\[
r_{11}=r,\text{ }r_{22}=1-r,\;r_{ij}=0\text{ otherwise,} 
\]
where r $\in \left[ 0,1\right] .$ For that pair $\left\{ \rho ,B\right\} $
eqs.$\left( \ref{03}\right) $ and $\left( \ref{20}\right) $ give 
\[
S_{12}=2r\left( 1-r\right) ,\text{ }\beta =r\xi _{1}+\left( 1-r\right) \xi
_{2}, 
\]
so that, for a fixed value of r (and therefore a fixed value of $S_{12}$ in
the interval $\left[ 0,1/2\right] )$ it is possible to choose $\xi _{1}$ and 
$\xi _{2}$ in such a way that $\beta $ takes any value in the interval $%
\left[ -2\sqrt{2\left( 1-S_{12}\right) },2\sqrt{2\left( 1-S_{12}\right) }%
\right] .$ In fact, for fixed r $\leq 1/2,$ $\xi _{1},\xi _{2}\geq 0,$ the
possible values of $\beta $ cover the interval $\left[ 2\sqrt{2}r,2\sqrt{%
2\left( 1-2r+2r^{2}\right) }\right] .$ For fixed r $\leq 1/2$ and $\xi
_{1}\geq 0,\xi _{2}\leq 0,$ the possible values of $\beta $ cover the
interval $\left[ -2\sqrt{2}\left( 1-r\right) ,2\sqrt{2}r\right] .$ Similar
analyses may be made for other choices, with the net results that the
possible values of $\xi _{1}$ and $\xi _{2}$ cover the whole interval $%
\left[ -2\sqrt{2\left( 1-2r+2r^{2}\right) },2\sqrt{2\left(
1-2r+2r^{2}\right) }\right] ,$ which proves that the image of $\Lambda _{2}$
is defined by the inequalities $\left( \ref{1a}\right) $ plus S$_{12}\leq
1/2.$

Hence the image of the union $\Lambda _{1}\cup \Lambda _{2}$ is defined by $%
\left( \ref{1a}\right) ,$ which proves the theorem. It follows at once:

\begin{corollary}
A sufficient condition for the fulfilment of all CHSH inequalities is that
the linear entropy of the state fulfils S$_{12}\geq \frac{1}{2}.$ For any
smaller value of S$_{12}$ there are states able to violate the inequalities.
\end{corollary}

The condition is not necessary, as may be trivially proved\cite{DJ}. For
instance, all pure states have S$_{12}$ = 0 but pure product states do not
violate a CHSH inequality.

For conditional entropies it is possible to prove similar results:

\begin{theorem}
In a two-qubit system, for any Bell operator, B, and any state, $\rho ,$ the
sum of the conditional linear entropies, $S_{2/1}+S_{1/2},$ of the state and
the parameter $\beta $\ lie within a region of the plane $\left\{ \beta
,S_{2/1}+S_{1/2}\right\} $ bounded by the inequalities 
\begin{equation}
S_{2/1}+S_{1/2}\geq -1,\;S_{2/1}+S_{1/2}\leq \frac{1}{2}-\frac{\beta ^{2}}{8}%
,\;S_{2/1}+S_{1/2}\leq 1-\frac{\beta ^{2}}{4},  \label{3}
\end{equation}
The region so defined is minimal in the sense that no smaller region of
compatibility exists.
\end{theorem}

In mathematical terms, this region is the image of the set $\Lambda $ in the
mapping $\left\{ \rho ,B\right\} $ $\rightarrow \left\{ \beta
,S_{1/2}+S_{1/2}\right\} .$

\textit{Proof}: In the set $\Lambda _{C},$ defined above, every density
matrix is diagonal in the basis of the associated Bell operator. This
implies that both reduced density matrices are multiple of the 2$\times $2
unit matrix. In fact, the reduced density matrices of a Bell state are both
multiple of the unit matrix, which implies that the same is true for any
mixture of Bell states. Therefore the sum of conditional entropies fulfils,
for these states, 
\[
S_{2/1}+S_{1/2}=2S_{12}-1. 
\]
If this is combined with $\left( \ref{1a}\right) $ we get the inequalities $%
\left( \ref{3}\right) ,$ which proves that the region which we are searching
for (the image of the set $\Lambda )$ contains the one defined by $\left( 
\ref{3}\right) $ (the image of the subset $\Lambda _{C}\subset \Lambda ).$
In particular, this means that any values $\left\{ \beta
,S_{1/2}+S_{1/2}\right\} $ within the region are the image of, at least, an
element of $\Lambda .$ Now we shall prove that both regions are in fact
identical, that is no element of $\Lambda $ has an image outside the region
defined by $\left( \ref{3}\right) .$ To do that we will prove these
inequalities for any element of $\Lambda .$

The first inequality $\left( \ref{3}\right) $ derives from the first of $%
\left( \ref{1a}\right) $ combined with the obvious one S$_{j}\leq 1/2,j=1,2.$
The proof of the second and third inequalities are similar to the proofs of
the second and third inequalities $\left( \ref{1a}\right) $ using, instead
of the density matrix $\rho ,$ the following matrix 
\begin{equation}
\rho ^{\prime }=\rho -\frac{1}{2}I_{1}\otimes \rho _{2}-\frac{1}{2}\rho
_{1}\otimes I_{2}+\frac{1}{2}I.  \label{4}
\end{equation}
This matrix $\rho ^{\prime }$ is Hermitean and has unit trace, but it is not
necessarily positive. Therefore it may not be a density matrix.
Nevertheless, the derivations do not require positivity of the matrix. In
fact, no additional condition besides being Hermitean and having unit trace
is required for the derivation of the second inequality. For the derivation
of the third we need also that the diagonal elements of the matrix in a
basis of Bell operators are positive. That this is true follows at once
taking into account that 
\[
\left\langle \psi \left| \rho ^{\prime }\right| \psi \right\rangle
=\left\langle \psi \left| \rho \right| \psi \right\rangle \geq 0, 
\]
for any Bell state $\psi ,$ as may be easily proved (see Appendix for the
explicit form of the Bell states). This completes the proof of the theorem.

\begin{corollary}
In a two-qubit system, a sufficient condition for the fulfilment of all CHSH
inequalities is that the sum of the conditional linear entropies of the
state is non-negative. For any negative value of S$_{1/2}$+S$_{2/1}$ there
are states able to violate the inequalities.
\end{corollary}

\section{Bell inequalities and von Neumann entropy}

In the following we shall derive similar results using, instead of the
linear entropy, the von Neumann entropy

\begin{equation}
S_{12}:=-Tr\left( \rho \ln \rho \right) ,S_{j}:=-Tr\left( \rho _{j}\ln \rho
_{j}\right) .  \label{6}
\end{equation}
Our problem is to find the image of the set $\Lambda ,$ of pairs $\left\{
\rho ,B\right\} ,$ in the mapping $N$ into the set of pairs $\left\{ \beta
,S_{12}\right\} \in \mathbf{R}^{2},$ $N$ being defined by eqs.$\left( \ref
{20}\right) $ and $\left( \ref{6}\right) .$

We begin searching for the image of the pairs $\left\{ \rho ,B\right\} $
when we fix the Bell operator B. We start considering the following family
of states, dependent on the parameter $\lambda \in \mathbf{R:}$ 
\begin{equation}
\rho =Z(\lambda )^{-1}\exp \left( \lambda B\right) ,\;Z(\lambda ):=Tr\exp
\left( \lambda B\right) ,  \label{8}
\end{equation}
the matrices $\rho $ so defined being Hermitean, positive and having unit
trace. It is straightforward to compute $\beta $ and $S_{12}$ from the
function $Z(\lambda )$ , that is 
\begin{equation}
\beta =\frac{d\ln Z}{d\lambda },\;S_{12}=-\frac{Tr\left\{ \exp \left(
\lambda B\right) \left[ \lambda B-\ln Tr\exp \left( \lambda B\right) \right]
\right\} }{Tr\exp \left( \lambda B\right) }=\ln Z-\lambda \beta .  \label{11}
\end{equation}

From the eigenvalues, $\left\{ \xi _{j}\right\} $, of the Bell operator (see 
$\left( \ref{1b}\right) )$ we get the function Z$\left( \lambda \right) $%
\[
Z\left( \lambda \right) =\exp \left( \lambda \xi _{1}\right) +\exp \left(
\lambda \xi _{2}\right) +\exp \left( -\lambda \xi _{2}\right) +\exp \left(
-\lambda \xi _{1}\right) =4\cosh \left( \lambda \mu \right) \cosh \left(
\lambda \nu \right) , 
\]
where $\mu =\frac{1}{2}\left( \xi _{1}+\xi _{2}\right) ,\nu =\frac{1}{2}%
\left( \xi _{1}-\xi _{2}\right) ,$ and we assume $\xi _{1}\geq \xi _{2}\geq
0.$ Hence it is straightforward to obtain $\beta $ and S$_{12}$ using eqs.$%
\left( \ref{11}\right) $ 
\begin{eqnarray}
\beta &=&\mu \tanh \left( \lambda \mu \right) +\nu \tanh \left( \lambda \nu
\right) ,  \nonumber \\
S_{12} &=&2\ln 2+\ln \cosh \left( \lambda \mu \right) +\ln \cosh \left(
\lambda \nu \right) -\lambda \beta .  \label{50}
\end{eqnarray}
For a fixed Bell operator (that is, fixed $\xi _{1}$ and $\xi _{2},$ or $\mu 
$ and $\nu )$ eqs.$\left( \ref{50}\right) $ provide the parametric equations
of a curve in the $\left\{ \beta ,S_{12}\right\} $ plane. The curve
contains, in particular, the points $\beta =0,S_{12}=2\ln 2$, for $\lambda
=0,$ and $\beta =\pm \xi _{1},S_{12}=0,$ for $\lambda \rightarrow \pm \infty
.$

Now we show that the image of all pairs $\left\{ \rho ,B\right\} ,$ with B
fixed, lie between the curve $\left( \ref{50}\right) $ and the straight line
S$_{12}$ = 0. The latter bound is obvious from the definition of entropy eq.$%
\left( \ref{6}\right) ,$ the former is proved as follows. We vary $\rho $
with the constraints 
\begin{equation}
Tr\rho =1\Rightarrow Tr\left[ \delta \rho \right] =0,\;Tr\left( \rho
B\right) =\beta \Rightarrow Tr\left[ B\delta \rho \right] =0,  \label{26}
\end{equation}
which leads to 
\[
\delta S_{12}=-Tr\left[ \left( \rho +\delta \rho \right) \ln \left( \rho
+\delta \rho \right) \right] +Tr\left( \rho \ln \rho \right) . 
\]
We may expand $\ln \left( \rho +\delta \rho \right) $ in powers of $\delta
\rho $ up to second order. The expansion is well defined because all integer
powers or $\rho ,$ eq.$\left( \ref{8}\right) ,$ either with positive or
negative exponent, are well defined. Also, to second order there is no
problem with the possible non-commutativity of the operators $\rho $ and $%
\delta \rho .$ Taking eqs.$\left( \ref{26}\right) $ into account we obtain
no term of first order in $\delta \rho ,$ and the second order term is 
\begin{equation}
\delta ^{(2}S_{12}=-\frac{1}{2}Tr\left[ \rho ^{-1}\delta \rho ^{2}\right]
\leq 0,  \label{334}
\end{equation}
which proves that a density operator of the form of eq.$\left( \ref{8}%
\right) $ makes S$_{12}$ a maximum for fixed B and $\beta .$

Now we consider the whole set $\Lambda $ of pairs $\left\{ \rho ,B\right\} .$
Its image will be the union of the images obtained for different Bell
operators B, that is the set of all points which lie between the curve S$%
_{12}$ = 0 and the highest of the curves $\left( \ref{50}\right) .$ The
highest curve corresponds to $\xi _{1}=2\sqrt{2},\xi _{2}=0,$ that is

\[
\beta =2\sqrt{2}\tanh x,\text{ \ }S_{12}=2\ln 2+2\ln \cosh x-2x\tanh
x,\;x\equiv \sqrt{2}\lambda . 
\]
From these equations it is possible to get explicitly S$_{12}$ as a function
of $\beta $ and we obtain 
\begin{equation}
S_{12}=2\ln 2-\left( 1+\beta ^{\prime }\right) \ln \left( 1+\beta ^{\prime
}\right) -\left( 1-\beta ^{\prime }\right) \ln \left( 1-\beta ^{\prime
}\right) ,\beta ^{\prime }=\frac{\beta }{2\sqrt{2}}.  \label{77}
\end{equation}
As a result of all this we have proved the following

\begin{theorem}
In a two-qubit system, for any Bell operator, B, and any state, $\rho ,$ the
von Neumann entropy of the state, S$_{12},$\ and the parameter $\beta $\ lie
within a region of the plane $\left\{ \beta ,S_{12}\right\} $ bounded by S$%
_{12}$ = 0 and eq.$\left( \ref{77}\right) .$ The region so defined is
minimal in the sense that no smaller region of compatibility exists.
\end{theorem}

In mathematical terms we may say that the image of the mapping \textit{N} is
the region between these two lines. Hence it follows at once

\begin{corollary}
If a two-qubit system is in a state with density matrix $\rho ,$ the
inequality 
\begin{equation}
S_{12}\geq 3\ln 2-\sqrt{2}\ln \left( \sqrt{2}+1\right) \simeq 0.833,
\label{49}
\end{equation}
where S$_{12}$ is the von Neumann entropy, is a sufficient condition for the
fulfillement of all CHSH inequalities. For any smaller value of S$_{12}$
there are states violating the inequalities.
\end{corollary}

Writing the density matrix $\left( \ref{8}\right) $ in the Bell state basis
it is easy to prove that the reduced density matrices are multiple of the
identity, that is 
\begin{equation}
\rho _{j}=\frac{1}{2}I_{j}\Rightarrow S_{1}=S_{2}=\ln 2.  \label{9}
\end{equation}
Thus we may prove

\begin{theorem}
In a two-qubit system, for any Bell operator, B, and any state, $\rho ,$ the
sum, S$_{1/2}+S_{2/1},$ of the conditional von Neumann entropies of the
state\ and the parameter $\beta $\ lie within a region of the plane $\left\{
\beta ,S_{12}\right\} $ bounded by 
\begin{eqnarray}
S_{1/2}+S_{2/1} &\leq &2\ln 2-2\left( 1+\beta ^{\prime }\right) \ln \left(
1+\beta ^{\prime }\right) -2\left( 1-\beta ^{\prime }\right) \ln \left(
1-\beta ^{\prime }\right) ,  \nonumber \\
S_{1/2}+S_{2/1} &\geq &-2\ln 2,\;\beta ^{\prime }=\frac{\beta }{2\sqrt{2}}
\label{10}
\end{eqnarray}
The region so defined is minimal in the sense that no smaller region of
compatibility exists.
\end{theorem}

\textit{Proof:} The second inequality follows trivially from $S_{12}\geq 0$
and $S_{j}\leq \ln 2.$ The first inequality is suggested by the previous
theorem. Here we show that it is in fact the upper bound by showing that,
for fixed B and $\beta $ (that is with the constraints $\left( \ref{26}%
\right) ),$ the variation 
\[
\delta \left( S_{2/1}+S_{1/2}\right) =2\delta S_{12}-\delta S_{1}-\delta
S_{2}
\]
is negative up to second order in $\delta \rho $. The second order term of $%
\delta S_{12}$ was obtained in $\left( \ref{334}\right) .$ Similarly we get 
\begin{equation}
\delta S_{j}=-Tr\left[ \left( \rho _{j}+\delta \rho _{j}\right) \ln \left(
\rho _{j}+\delta \rho _{j}\right) \right] +Tr\left( \rho _{j}\ln \rho
_{j}\right) =-Tr\left[ \delta \rho _{j}^{2}\right] +O\left( \delta \rho
_{j}^{3}\right) ,  \label{25}
\end{equation}
where 
\[
\rho _{1}=Tr_{2}\rho ,\delta \rho _{1}=Tr_{2}\left( \delta \rho \right) ,
\]
and similar for $\rho _{2}$ and $\delta \rho _{2}$. In the second eq.$\left( 
\ref{25}\right) $ we have taken into account eqs$.\left( \ref{26}\right) $
and $\left( \ref{9}\right) ,$ the latter implying $\rho _{j}^{-1}=2I_{j}$
and ln$\rho _{j}=-\ln 2$ $I_{j}$. Hence, using eqs.$\left( \ref{334}\right) $
and $\left( \ref{25}\right) $, we get 
\begin{equation}
\delta \left( S_{2/1}+S_{1/2}\right) =Tr\left[ \delta \rho _{1}^{2}\right]
+Tr\left[ \delta \rho _{2}^{2}\right] -Tr\left[ \rho ^{-1}\delta \rho
^{2}\right] +O\left( \delta \rho ^{3}\right) .  \label{27}
\end{equation}
Now we shall compare $\left( \ref{25}\right) $ with $\left( \ref{334}\right) 
$ using the obvious inequality 
\[
\left( \delta \rho -\frac{1}{2}I_{1}\otimes \delta \rho _{2}-\frac{1}{2}%
\delta \rho _{1}\otimes I_{2}+\frac{1}{2}I\right) ^{2}\geq 0,
\]
which gives, after some algebra with the first eq.$\left( \ref{26}\right) $
taken into account, 
\[
Tr\left[ \delta \rho _{1}^{2}\right] +Tr\left[ \delta \rho _{2}^{2}\right]
\leq Tr\left[ \delta \rho ^{2}\right] .
\]
Hence eq.$\left( \ref{27}\right) $ gives, to second order in $\delta \rho ,$ 
\[
\delta ^{(2}\left( S_{2/1}+S_{1/2}\right) \leq Tr\left[ \delta \rho
^{2}\right] -Tr\left[ \rho ^{-1}\delta \rho ^{2}\right] .
\]
The right hand side may be calculated in a basis of Bell states and we
obtain 
\begin{eqnarray*}
\delta ^{(2}\left( S_{2/1}+S_{1/2}\right)  &\leq &\sum_{k=1}^{4}\langle \psi
_{k}\mid \delta \rho ^{2}\left[ 1-\rho ^{-1}\right] \mid \psi _{k}\rangle  \\
&=&\sum_{k=1}^{4}\langle \psi _{k}\mid \delta \rho ^{2}\mid \psi _{k}\rangle
\left[ 1-Z\left( \lambda \right) \exp \left( -\lambda \xi _{k}\right)
\right] ,
\end{eqnarray*}
where we have labelled $\mid \psi _{k}\rangle $ the Bell states and $\xi _{k}
$ the corresponding eigenvalues, $Z\left( \lambda \right) $ being given by
eq.$\left( \ref{8}\right) .$ We see that the right hand side is negative if
the following inequality holds for every $k$ 
\[
Z\left( \lambda \right) \exp \left( -\lambda \xi _{k}\right) >1,
\]
which is trivially true because $Z\left( \lambda \right) $ is the sum of
four positive numbers, one of them being $\exp \left( \lambda \xi
_{k}\right) .$This shows that $S_{2/1}+S_{1/2}$ presents a maximum, for
given B and $\beta ,$ if $\rho $ is given by eq.$\left( \ref{8}\right) .$ 
The rest of the proof parallels the one of the previous theorem. It involves
showing that the first eq.$\left( \ref{10}\right) $ is the highest of the
curves $S_{2/1}+S_{1/2}$ versus $\beta ,$ one curve for each B.

\begin{corollary}
If a two-qubit system is in a state with density matrix $\rho ,$ the
inequality 
\[
S_{2/1}+S_{1/2}\geq 4\ln 2-2\sqrt{2}\ln \left( \sqrt{2}+1\right) \simeq
0.280,
\]
in terms of von Neumann entropy, is a sufficient condition for the
fulfillement of all CHSH inequalities. For any smaller value, there are
states violating the inequalities.
\end{corollary}

It is interesting that, according to this theorem, the second implication $%
\left( \ref{implications}\right) $ does not hold true in the case of the von
Neumann entropy. In fact, for $S_{2/1}+S_{1/2}=0$ it is possible to get
values for the parameter $\beta $ as high as 2.206, which may be easily
derived from the first eq.$\left( \ref{10}\right) .$

\section{Entropy and local hidden variables}

I shall finish with a comment about how specific for the CHSH inequalities
are the results here presented, that is whether they may be extended to
other Bell inequalities (i.e. inequalities characteristic of local hidden
variables (LHV) models ). The question, stated more generally, is whether
the entropy inequalities considered in the previous theorems are sufficient
for the existence of LHV models. The answer seems to be negative, although a
more detailed study is necessary. In fact, it is known that the CHSH
inequalities are necessary conditions for the existence of LHV theories, but
they are not sufficient. It has been proved that, chosen four observables a$%
_{1}$, a$_{2}$, b$_{1}$, b$_{2}$ as in eq.$\left( \ref{5}\right) ,$ the
fulfillement of the four CHSH inequalities obtained by changing the place of
the minus sign, is a sufficient condition for the existence of a LHV model
involving these four observables\cite{fine}, but there are counterexamples
proving that the condition is not sufficient for more than four\cite{mermin}.

\section{Appendix}

For the sake of clarity I present here a short rederivation of some
properties of the Bell operator (see the paper by Braunstein et al.\cite
{revzen}.)

The square of the Bell operator $\left( \ref{18}\right) $ may be written,
taking into account that the square of any of the operators a$_{1}$, a$_{2}$%
, b$_{1}$, or b$_{2}$ is the unit operator in the corresponding Hilbert
space, 
\[
B^{2}=4I_{1}\otimes I_{2}-\left[ a_{1},b_{1}\right] \otimes \left[
a_{2},b_{2}\right] . 
\]
Now we remember that any operator, $a,$ in a two-dimensional space having
eigenvalues $\pm 1$ may be written in the form 
\[
a=\mathbf{a\cdot \sigma ,} 
\]
where \textbf{a} is a unit vector in ordinary, three-dimensional, space and $%
\mathbf{\sigma }$ the vector of the Pauli matrices. Thus we may write 
\begin{eqnarray*}
B^{2} &=&4I_{1}\otimes I_{2}+4(\mathbf{a}_{1}\times \mathbf{b}_{1})\cdot 
\mathbf{\sigma }_{1}\otimes (\mathbf{a}_{2}\times \mathbf{b}_{2})\cdot 
\mathbf{\sigma }_{2} \\
&\equiv &4I_{1}\otimes I_{2}+4\left| \mathbf{a}_{1}\times \mathbf{b}%
_{1}\right| \left| \mathbf{a}_{2}\times \mathbf{b}_{2}\right| \sigma
_{1z}\otimes \sigma _{2z},
\end{eqnarray*}
where the last expression corresponds to taking reference frames with the z
axis in the direction $\mathbf{a}_{1}\times \mathbf{b}_{1}$ ($\mathbf{a}%
_{2}\times \mathbf{b}_{2})$ for the first (second) particle. From the latter
representation it is easy to see that B$^{2}$ possesses eigenvectors which
may be represented, with an obvious notation,

$\mid \uparrow \uparrow \rangle $ and $\mid \downarrow \downarrow \rangle $
both with eigenvalue ($\xi _{1})^{2}=(\xi _{4})^{2}=$ $4+4\left| \mathbf{a}%
_{1}\times \mathbf{b}_{1}\right| \left| \mathbf{a}_{2}\times \mathbf{b}%
_{2}\right| ,$

$\mid \uparrow \downarrow \rangle $ and $\mid \downarrow \uparrow \rangle $
both with eigenvalue ($\xi _{2})^{2}=(\xi _{3})^{2}=$ $4-4\left| \mathbf{a}%
_{1}\times \mathbf{b}_{1}\right| \left| \mathbf{a}_{2}\times \mathbf{b}%
_{2}\right| .$

Hence eqs.$\left( \ref{1b}\right) $ and $\left( \ref{19}\right) $ follow
without difficulty. It is easy to see that the four eigenvectors of B are $%
\frac{1}{\sqrt{2}}\left( \mid \uparrow \uparrow \rangle \pm \mid \downarrow
\downarrow \rangle \right) ,\frac{1}{\sqrt{2}}\left( \mid \uparrow
\downarrow \rangle \pm \mid \downarrow \uparrow \rangle \right) ,$ usually
called Bell states.

\textbf{Acknowledgement}. I thank L. Jak\'{o}bczyk\cite{J} for pointing out
an the error in a previous version of this paper\cite{santos}.

\smallskip

\end{document}